\documentclass[12pt]{article}
\usepackage{graphicx}
\usepackage{hyperref}

\newcommand\pubnumber{}
\newcommand\pubdate{\today}

\def\napoli{The Oskar Klein Centre for Cosmoparticle Physics,\\
Department of Physics, Stockholm University, SE-106 91 Stockholm, Sweden}
\def\support{\footnote{Work based on Gerbino \& Lattanzi, \textit{Status of neutrino properties and future prospects - Cosmological and astrophysical constraints}, Front.in Phys. 5 (2018) 70~\cite{Lattanzi:2017ubx}. Slides available at \href{https://indico.ph.qmul.ac.uk/indico/getFile.py/access?contribId=25&resId=0&materialId=slides&confId=170}{NuPhys2017 website}.}}

\def\Title#1{\begin{center} {\Large #1 } \end{center}}
\def\Author#1{\begin{center}{ \sc #1} \end{center}}
\def\Address#1{\begin{center}{ \it #1} \end{center}}

\newcommand\pubblock{\rightline{\begin{tabular}{l} \pubnumber\\
         \pubdate  \end{tabular}}}
\newenvironment{Abstract}{\begin{quotation}  }{\end{quotation}}
\newenvironment{Presented}{\begin{quotation} \begin{center} 
             PRESENTED AT\end{center}\bigskip 
      \begin{center}\begin{large}}{\end{large}\end{center} \end{quotation}}
\def\Acknowledgements{\bigskip  \bigskip \begin{center} \begin{large}
             \bf ACKNOWLEDGEMENTS \end{large}\end{center}}

\def\beq{\begin{equation}}
\def\eeq#1{\label{#1}\end{equation}}
\def\eeqn{\end{equation}}

\def\beqa{\begin{eqnarray}}
\def\eeqa#1{\label{#1}\end{eqnarray}}
\def\eeqan{\end{eqnarray}}

\let\bar=\overbar

\def\Dslash{\not{\hbox{\kern-4pt $D$}}}
\def\dslash{\not{\hbox{\kern-2pt $\del$}}}

\def\msb{{\bar{\ssstyle M \kern -1pt S}}}

\begin{document}
\begin{titlepage}
\pubblock

\vfill
\Title{Neutrino properties from cosmology}
\vfill
\Author{Martina Gerbino\support}
\Address{\napoli}
\vfill
\begin{Abstract}
Precision cosmology enables to test fundamental physics, including neutrino properties, with unprecedented accuracy. In this work, I review the basics of neutrino cosmology. I briefly describe how neutrinos affect cosmological observables, such as anisotropies in the cosmic microwave background and fluctuations in the matter field. I show current constraints and projected sensitivities on the sum of the neutrino masses $\Sigma m_\nu$ and on the number of relativistic species $N_\mathrm{eff}$ from a selection of cosmological data. I finally comment about the implications of those bounds for neutrino physics.
\end{Abstract}
\vfill
\begin{Presented}
NuPhys2017, Prospects in Neutrino Physics\\
Barbican Centre, London, UK,  December 20--22, 2017
\end{Presented}
\vfill
\end{titlepage}
\def\thefootnote{\fnsymbol{footnote}}
\setcounter{footnote}{0}

\section{Introduction}

The observation of neutrino flavour oscillations provides the experimental evidence that neutrinos are massive particles (see e.g.~\cite{deSalas:2017kay} for a recent global fit of neutrino oscillation parameters). This represents a strong evidence of physics beyond the standard model (SM), since it is not possible to build a neutrino mass term from the set of SM particles. 

Cosmology is a preferred arena for the investigation of neutrino properties. Neutrinos deeply affect the background evolution of the Universe, as well as the evolution of cosmological perturbations and structure formation. Therefore, the combination of various cosmological probes, such as anisotropies in the cosmic microwave background (CMB) and fluctuations in the matter field, provides stringent constraints on several neutrino properties. These include the sum of the neutrino masses $\Sigma m_\nu$ and the number of relativistic species $N_\mathrm{eff}$ in the early Universe. 

In this work, I will briefly review the basics of neutrino cosmology in Sec.~\ref{sec:nucosmo} (for detailed reviews, see e.g.~\cite{Lattanzi:2017ubx,Lesgourgues:2006nd,Dolgov:2002wy}). I will mostly focus on the three families of active neutrinos, although I will briefly comment about cosmological constraints on additional relativistic species later on. In Sec.~\ref{sec:limits}, I will summarize current bounds on neutrino properties from a compilation of cosmological observables, and comment about expected sensitivities from future cosmological observatories. Conclusions can be found in Sec.~\ref{sec:conclusion}.

\section{Basics of Neutrino Cosmology}\label{sec:nucosmo}
\subsection{The Cosmic Neutrino Background}\label{subsec:cnb}
The existence of a cosmological background of relic neutrinos is a prediction of the standard cosmological model. In the early Universe, these relic neutrinos are in thermal equilibrium with the rest of the primordial plasma via weak interactions, as long as the weak-interaction rate $\Gamma$ is higher than the Hubble expansion rate $H$. The two rates are fixed at any given temperature $T$, so that it is easy to determine the epoch of neutrino decoupling by setting $\Gamma(T_\mathrm{\nu,dec})=H(T_\mathrm{\nu,dec})$. The weak interaction rate is $\Gamma\propto G_F^2 T^5$, where $G_F$ is the Fermi constant. The expansion rate at the time of neutrino decoupling is basically set by the energy density of radiation, $H\propto T^2$. It follows that neutrinos decouple from the rest of the primordial plasma for $T_\mathrm{\nu,dec}\simeq 1\,\mathrm{MeV}$, therefore while still relativistic. The neutrino temperature after decoupling scales as the inverse of the scale factor $T_\nu\propto 1/a$, while the common temperature of the plasma continues to evolve as $T\propto 1/(a g_{*}^{1/3})$ where $g_{*}$ is the effective number of degrees of freedom. 

Soon after neutrino decoupling, the temperature of the plasma drops below the threshold for $e^+ e^-$ pair production. The annihilation of $e^+ e^-$ transfers entropy to the rest of the plasma, i.e. CMB photons. Neutrinos, that are already decoupled, do not benefit from the entropy transfer. As a result, the CMB temperature today is slightly higher than the relic neutrino temperature. Entropy conservation allows to compute the ratio of the two temperatures today as $T_\gamma/T_\nu=(11/4)^{1/3}$. Precise measurements of the COBE satellite sets the CMB temperature today to be  $T_\gamma=(2.725\pm0.002)\,\mathrm{K}$~\cite{Mather:1998gm,Fixsen:1996nj}. Therefore, the neutrino temperature can be estimated to be $T_\nu\simeq 1.95\,\mathrm{K}\simeq 1.68\cdot 10^{-4}\,\mathrm{eV}$. 

Neutrinos are fully relativistic particles in the early Universe, thus they contribute as radiation to the energy density of the Universe. Their contribution to the radiation density after $e^+ e^-$ annihilation can be parametrized in terms of the phenomenological parameter $N_\mathrm{eff}$:
\begin{equation}\label{eq:neff}
\rho_\mathrm{rad}=\rho_\gamma\left[1+\frac{7}{8}\left(\frac{4}{11}\right)^{\frac{4}{3}}N_\mathrm{eff}\right].
\end{equation}

The parameter $N_\mathrm{eff}$ is defined as the difference between the total radiation density and the CMB energy density, normalized to the energy density of an individual neutrino species. Therefore, it accounts for \textit{any} relativistic species which might be present at early times. In the standard cosmological model, only the three families of active neutrinos are present and $N_\mathrm{eff}=3.045$~\cite{deSalas:2016ztq,Mangano:2005cc,Dolgov:2002wy}. The extra contribution $N_\mathrm{eff}-3=0.045$ with respect to the three families of active neutrinos is an exact result of the complete treatment of neutrino decoupling, which takes into account non-instantaneous decoupling, finite temperature QED radiative corrections and flavor oscillations. 

On the other hand, we know that at least two families of neutrinos are non-relativistic today, as we observe two non-vanishing mass splittings from oscillation experiments~\cite{deSalas:2017kay}. The redshift of transition to the non-relativistic regime is roughly given by $z_\mathrm{NR}\simeq1900 (m_\nu/\mathrm{eV})$, where $m_\nu$ is the mass of an individual neutrino mass state. At redshift $z<z_\mathrm{NR}$, neutrinos contribute as a matter component to the energy density of the Universe. Their contribution is expressed in terms of the sum of the neutrino masses:
\begin{equation}\label{eq:numass}
\Omega_\nu h^2=\frac{\Sigma m_\nu}{93.14\,\mathrm{eV}},
\end{equation}
where $\Omega_\nu=\rho_{\nu,0}/\rho_{c,0}$ is neutrino density parameter today, expressed as the ratio between the neutrino energy density $\rho_{\nu,0}$ and the critical density $\rho_{c,0}=3H_0^2/(8\pi G)$ today, and $h$ is the reduced Hubble constant $H_0=100 h\,\mathrm{km\,s^{-1}\,Mpc^{-1}}$.

\subsection{How neutrinos affect cosmology}\label{subsec:impact}
A direct observation of the cosmic neutrino background is challenging, although experimental efforts are devoted to this task~\cite{Betts:2013uya}. Nevertheless, the peculiar imprints that relic neutrinos have on the evolution of cosmological observables are extremely useful to indirectly probe neutrino properties. 
We can distinguish two generic classes of effects that neutrinos have on cosmological probes, namely effects on the background evolution of the Universe and effects on the evolution of cosmological perturbations. In this section, we review briefly both classes of effects. We will see in Sec.~\ref{subsec:probes} how these effects are actually imprinted in the cosmological observables.

Let us focus on the first class, i.e. the background effects. Neutrinos make up a significant fraction of the energy density of the Universe, and we have seen in Sec.~\ref{subsec:cnb} how this contribution can be expressed at different epochs. Therefore, neutrinos contribute to setting the expansion rate $H(z)$ at any given redshift:
\begin{equation}\label{eq:Hz}
H(z)^2=H_0^2\left[(\Omega_c+\Omega_b)(1+z)^3+\Omega_\gamma(1+z)^4+\Omega_\Lambda+\frac{\rho_\nu(z)}{\rho_{\mathrm{c},0}}\right],
\end{equation}
where $\Omega_c+\Omega_b$ are the baryonic and cold dark matter density parameters, $\Omega_\gamma$ is the photon density parameter, and $\Omega_\Lambda$ is the dark energy density parameter in the form of a cosmological constant (all evaluated today). The neutrino energy density $\rho_\nu(z)$ scales differently with the redshift, depending whether neutrinos account for either radiation or matter. A change in $\rho_\nu(z)$ while keeping fixed the other parameters corresponds to a change in $H(z)$. This in turn is reflected into changes in the angles subtended by some peculiar scales in cosmology and imprinted in cosmological observables. These angular scales are usually defined as the ratio between the physical scale of interest $r$ and the angular diameter distance $D_A(z)=\int_0^{z} \mathrm{d}z^{'}/H(z^{'})$. Important examples are $\theta_s(z_\mathrm{rec})$, the angular size of the sound horizon at the time of hydrogen recombination $z_\mathrm{rec}\simeq1100$~\cite{Ade:2015xua}; $\theta_s(z_\mathrm{drag})$,  the angular size of the sound horizon at the time of baryon decoupling from CMB photons $z_\mathrm{drag}\simeq1060$~\cite{Ade:2015xua}; and $\theta_s(z_\mathrm{d})$ the angular scale of the damping induced on CMB photons by scattering off free electrons prior to recombination (Silk damping).

In addition to a change in $H(z)$, neutrinos can modify the relative amount of matter and radiation. The main consequence of this fact is that the redshift $z_\mathrm{eq}$ at which the matter energy density equates the radiation energy density (epoch of matter-radiation equality) can be shifted either backward or forward in time:
\begin{equation}\label{eq:zeq}
z_\mathrm{eq}=\frac{\Omega_c+\Omega_b}{\Omega_\gamma \left[1+\frac{7}{8}\left(\frac{4}{11}\right)^{\frac{4}{3}}N_\mathrm{eff}\right]}
\end{equation}
In the standard cosmological model, neutrinos are fully relativistic at $z_\mathrm{eq}\simeq3400$~\cite{Ade:2015xua} and $N_\mathrm{eff}=3.045$. Therefore, the denominator of Eq.~\ref{eq:zeq} is fixed. However, a change in $\Sigma m_\nu$ modifies the total matter density at late times. In a flat cosmological model, this implies that $\Omega_c+\Omega_b$ has to be modified accordingly to satisfy the flatness constraint, and the numerator of Eq.~\ref{eq:zeq} is modified as well. Of course, $z_\mathrm{eq}$ can be also altered by a change in $N_\mathrm{eff}$, as it is the case in some non-standard cosmological scenarios (e.g. presence of extra-relativistic species, low-reheating models).

We now move to consider the second class of neutrino effects on cosmological observables, namely effects at the level of cosmological perturbations in the gravitational potentials and the other cosmological components. The first effect we consider is somehow connected to the shift in $z_\mathrm{eq}$. Anisotropies in the CMB carry the imprint of time-variation of gravitational potentials along the line of sight from the time of emission of the CMB to the observer (integrated Sachs-Wolfe effect, ISW). Gravitational potentials are constant during the matter-dominated epoch, and exhibit time-variation during the radiation-dominated epoch (early ISW) and during the current dark-energy epoch (late ISW). A change in $z_\mathrm{eq}$ can therefore either enhance or inhibit the early ISW effect.

Before $z_\mathrm{rec}$, photons and baryons are tightly coupled and this photon-baryon fluid undergo acoustic oscillations due to the counteraction of gravity and radiation pressure. The phase of these oscillations can be altered by free-streaming particles propagating faster than the sound speed of the fluid, such as relativistic neutrinos~\cite{Baumann:2017lmt,Baumann:2015rya}. Therefore, a change in $N_\mathrm{eff}$ can modify the phase of the acoustic oscillations. 

The evolution of matter perturbations is strongly affected by massive neutrinos, see e.g.~\cite{Hu:1997mj} for one of the pivotal studies on the use of large-scale-structure observations to constrain massive neutrinos. In this sense, it is useful to define a free-streaming scale 
\begin{equation}\label{eq:kfs}
k_\mathrm{fs}=0.018\,\Omega_m^{1/2}\left(\frac{m_\nu}{1\,\mathrm{eV}}\right)^{1/2} h\,\mathrm{Mpc^{-1}}
\end{equation}
that corresponds to the size of the sound horizon at the time of neutrino non-relativistic transition. Below this scale ($k\gg k_\mathrm{fs}$), neutrinos have a large thermal velocity and do not contribute to the formation of structures, effectively washing out perturbations. Above this scale ($k\ll k_\mathrm{fs}$), neutrinos behave as a matter component and contribute to clustering. This different scale-dependent behaviour has two main consequences. 

First, the growth of matter perturbations is delayed below $k_\mathrm{fs}$. This happens because the volume contained within a linear scale smaller than the free-streaming length resembles a mixed matter-radiation Universe, as opposed to the purely matter-dominated case of $k\ll k_\mathrm{fs}$. In a purely matter-dominated Universe, perturbations $\delta$ in the matter fluid evolve proportionally to the scale factor, therefore $\delta_m(k\ll k_\mathrm{fs})\propto a$. In a mixed Universe, where massive neutrinos make up a fraction $f_\nu$ of the total matter density, the matter perturbations evolve slowly, as $\delta_m(k\gg k_\mathrm{fs})\propto a^{1-3/5f_\nu}$.
Secondly, the clustering of structures is also scale-dependent, with a depletion of power at small scales. This is due to the fact that free-streaming neutrinos can escape the potential wells in which matter tends to cluster. In contrast, neutrinos effectively resemble a cold dark matter component at the largest scales. 

In the next section, we will see how the effects detailed above modify the cosmological observables.

\subsection{Cosmological observables}\label{subsec:probes}
The main cosmological probes are the CMB~\cite{Hu:2001bc} and the distribution of matter in the Universe (see e.g.~\cite{Weinberg:2012es} for a review on methods for measuring the expansion history and the growth of structures). We can map the anisotropies in the CMB (one intensity map $T$, and two polarization maps $E$ and $B$) and the fluctuations in the matter fields. The standard model assumption of Gaussian perturbations allows to encode all the information contained in those maps in the two-point correlation functions of the fields, or equivalently in their power spectra. Thus, the relevant observable for the CMB is the power spectrum $C_\ell^{XY}$. The multipole $\ell$ is the inverse of the angular separation between two directions in the sky, and $X,Y$ can be any of $T,E,B$. The relevant observable for the matter field is the power spectrum $P_m(k,z)$, where $k$ is the Fourier wavenumber and $z$ is the redshift\footnote{In addition to the clustering of matter at cosmological scales, there are other probes of the large-scale structure of Universe. For example, the observation of the variation of the number of galaxy clusters of a certain mass $M$ with redshift $dN(z,M)/dz$ (cluster counts)~\cite{Carlstrom:2002na}, the observation of the weak gravitational lensing~\cite{Kilbinger:2014cea}, and the reconstruction of the lensing power spectrum (see e.g.~\cite{Hu:2001kj}) are powerful probes of the low-redshift Universe. In this work, we focus on clustering for pedagogical purposes. The interested reader can refer e.g. to~\cite{Lattanzi:2017ubx} for an extensive description of other cosmological and astrophysical probes.}. In practice, we do not observe the total matter spectrum (cold dark matter and baryonic matter). In fact, we have observational access to certain tracers of the underlying matter field, for example the clustering of galaxies up to a certain redshift. The spectrum of the tracer provides a biased estimation of the total matter power spectrum, $P_\mathrm{tracer}=b^2_\mathrm{tracer}(k,z)P_m$. The imperfect knowledge of the bias $b_\mathrm{tracer}$, and especially its scale-dependence, might limit the constraining power of the full shape of the matter power spectrum as a cosmological probe (see e.g.~\cite{Giusarma:2018jei} for a recent work). 

The exact shape of the CMB and matter power spectra depends on the physics outlined in Sec.~\ref{subsec:impact}. The acoustic oscillations propagating in the baryon-photon fluid have produced the characteristic sequence of peaks and throughs in the CMB power spectra. The series of acoustic peaks in the small-scale region of the matter power spectrum, the so-called Baryon Acoustic Oscillations (BAO), have also the same origin. The position of the acoustic peaks, both in the CMB spectra and in the matter spectrum, depends strongly on the phase of the acoustic oscillations. 

The epoch of matter-radiation equality precisely determines the amplitude of the first peak in the CMB temperature power spectrum through the early ISW, and sets the position of the ``turning point'' (the main peak) of the matter power spectrum. The position of the first peak in the CMB TT spectrum precisely constrains the angular scale of the sound horizon $\theta_s$, whereas the angular scale of the Silk damping $\theta_d$ identifies the multipole range which is suppressed in power. In a similar way, the free-streaming scale $k_\mathrm{fs}$ determines the scale at which the matter power begins to be suppressed with respect to larger scales. The effect of massive neutrinos on the evolution of structures is also imprinted indirectly on the small-scale region of the CMB spectra, through the smearing of the acoustic peaks due to gravitational lensing~\cite{Lewis:2006fu}. 

From what said above, it is clear that one could constrain with high accuracy $N_\mathrm{eff}$ and $\Sigma m_\nu$ by looking at the features they induce in the CMB and matter power spectra. In practice, other cosmological parameters concur at shaping the same spectra, and there is the possibility to compensate the effect of a certain parameter by adjusting another. For example, in Eq.~\ref{eq:Hz}, an increase in $\rho_\nu$ can be compensated by a decrease in $\Omega_\Lambda$ in order to keep the expansion rate fixed. To overcome this limitation, the combination of multiple cosmological probes is particularly powerful. Each observable is sensitive to a specific combination of parameters, so that the cross correlation of multiple probes allows to break parameter degeneracies and to get tighter constraints.

Before moving to the next section, let us briefly comment about the possibility to go beyond bounds on $\Sigma m_\nu$. From Sec.~\ref{subsec:impact}, it is clear that cosmological probes are mostly sensitive to $\Sigma m_\nu$. In principle, there exist a free-streaming scale for each neutrino mass state. Since the mass states have slightly different masses, there are three different free-streaming scales, and one might be able to see the effects of each scale imprinted on cosmological probes. This would allow to discriminate between the two different neutrino hierarchies, namely the normal hierarchy ($m_1\simeq m_2\ll m_3$) and the inverted hierarchy ($m_3\ll m_1\simeq m_2$). In practice, the sensitivity of current data -- and likely future data -- is not enough to identify the effects of the individual mass states~\cite{Gerbino:2016ehw}. For all the practical purposes in cosmological studies, neutrinos are assumed to be degenerate in mass ($m_{1,2,3}=\Sigma m_\nu /3$) and it can be shown that this is a valid approximation of the real mass spectrum~\cite{Lesgourgues:2004ps}. The most viable solution to discriminate between the two neutrino hierarchies with cosmological data is to reach the sensitivity needed to exclude at high statistical significance the lowest value of $\Sigma m_\nu$ allowed by neutrino oscillation data for the inverted hierarchy ($\Sigma m_{\nu,\mathrm{min}}=0.1\,\mathrm{eV}$).

\section{Constraints on $N_\mathrm{eff}$ and $\Sigma m_\nu$}\label{sec:limits}
The tightest constraints on $\Sigma m_\nu$ from a single observable come from the measurements of the CMB anisotropies from the Planck satellite~\cite{Aghanim:2015xee}. In the framework of the standard cosmological model $\mathrm{\Lambda CDM}$ with the addition of massive neutrinos, the full set of temperature data (TT) with the addition of the large-scale polarization (lowP) gives $\Sigma m_\nu<0.72\,\mathrm{eV}$ at 95\% confidence level (CL)~\cite{Ade:2015xua}. This limit is already tighter than the bounds on the neutrino mass scale obtained with laboratory searches, such as limits from $\beta$-decay experiments~\cite{Aseev:2011dq,Kraus:2004zw} and neutrinoless double-$\beta$-decay experiments~\cite{0n2b}. The inclusion of large-scale structure data in the form of BAO measurements\footnote{See Ref.~\cite{Ade:2015xua} for a complete list of the BAO measurements concurring to the bounds reported here.} improves the limit from Planck alone, and gives $\Sigma m_\nu<0.25\,\mathrm{eV}$ at 95\% CL~\cite{Ade:2015xua}. Further improvements can be obtained with the combination of other astrophysical probes, such as Supernovae-Ia data and direct measurements of the Hubble constant, or with more aggressive analysis (see e.g. the discussion in~\cite{Vagnozzi:2017ovm} and references therein). The interested reader can find an extensive list of current bounds in~\cite{Lattanzi:2017ubx} and references therein. For the purpose of this work, it is sufficient to show that CMB data in combination with large-scale-structure measurements are a powerful tool to constrain the neutrino mass scale. Indeed, the current bounds are approaching the non-degenerate region of the neutrino mass spectrum, where for a given value of the lightest mass state the two neutrino hierarchies would predict a significantly different value of $\Sigma m_\nu$. The current figures are not far from the threshold of $\Sigma m_{\nu,\mathrm{min}}=0.1\,\mathrm{eV}$ mentioned at the end of Sec.~\ref{subsec:impact} and to the possibility to rule out the inverted hierarchy. 

The projected sensitivity on $\Sigma m_\nu$ from the next generation of cosmological probes, such as CMB Stage-IV and the DESI large-scale-structure survey, is $\sigma(\Sigma m_\nu)\simeq 0.015\,\mathrm{eV}$~\cite{Abazajian:2016yjj}. This sensitivity would allow a $3\sigma$ detection of the minimal mass allowed by neutrino oscillation data in the normal hierarchy scenario ($\Sigma m_{\nu,\mathrm{min}}=0.06\,\mathrm{eV}$). Provided that the real value of $\Sigma m_\nu$ is much lower than $\Sigma m_{\nu,\mathrm{min}}=0.1\,\mathrm{eV}$, the same sensitivity would also reject the inverted hierarchy at high statistical significance.

A final note on the limits on $\Sigma m_\nu$. When derived in more extended cosmological scenarios, such as models with arbitrary curvature and/or dark energy, the bounds on $\Sigma m_\nu$ can degrade due to the aforementioned degeneracies with the other cosmological parameters. For example, the limits on $\Sigma m_\nu$ from Planck+BAO in the context of $\mathrm{\Lambda CDM}$ with massive neutrinos and arbitrary curvature are 30\% broader than the corresponding bounds derived in a flat $\mathrm{\Lambda CDM}$ with massive neutrinos~\cite{Ade:2015xua}. Of course, the inclusion of additional cosmological data can help break degeneracies and improve the limits. The increased sensitivity of future cosmological surveys will also alleviate this issue and reduce parameter degeneracies, see e.g.~\cite{Archidiacono:2016lnv}.

Moving to the current constraints on $N_\mathrm{eff}$, Planck TT+lowP provides a 68\% CL bound of $N_\mathrm{eff}=3.13\pm0.32$~\cite{Ade:2015xua}. The inclusion of BAO data improves this bound to $N_\mathrm{eff}=3.15\pm0.23$~\cite{Ade:2015xua}. These limits imply that cosmology prefers a value of $N_\mathrm{eff}$ in agreement with the predictions of the standard cosmological model. The presence of an additional fully-thermalized species which would contribute to $N_\mathrm{eff}$ as $\Delta N_\mathrm{eff}=1$ is excluded at $>3\sigma$ level. This is the case of an additional light sterile neutrino with the mass and mixing angle necessary to solve reactor anomalies. When cosmological data are interpreted in the context of $\mathrm{\Lambda CDM}$ with arbitrary $N_\mathrm{eff}$ and light sterile neutrinos, the joint 95\% CL constraints on $N_\mathrm{eff}$ and the effective sterile neutrino mass $m_\mathrm{eff}=(94.1\,\Omega_{\nu,sterile}h^2)\,\mathrm{eV}$ are: $N_\mathrm{eff}<3.7$, $m_\mathrm{eff}<0.38\,\mathrm{eV}$~\cite{Ade:2015xua}.

The Stage-IV generation of cosmological surveys will reach the sensitivity of $\sigma(N_\mathrm{eff})=0.027$~\cite{Abazajian:2016yjj}. This is an important theoretical threshold, because it corresponds to the contribution to $N_\mathrm{eff}$ of a Goldstone boson decoupling before the QCD phase transition~\cite{Baumann:2016wac}. Moreover, the expected sensitivity on $N_\mathrm{eff}$ will allow to test with unprecedented accuracy the physics of non-instantaneous neutrino decoupling. In fact, the future sensitivity will allow to distinguish between $N_\mathrm{eff}=3$ and $N_\mathrm{eff}=3.045$ at $>1\sigma$ level.  

\section{Conclusion}\label{sec:conclusion}
We are currently in the epoch of precision cosmology. The current sensitivity reached by cosmological experiments as well as the forecasted performances of the next generation of cosmological surveys allow to test fundamental physics with unprecedented accuracy. Cosmology already provides the tightest bounds on the sum of the neutrino masses $\Sigma m_\nu$. The combination of CMB data from Planck with BAO measurements gives $\Sigma m_\nu<0.25\,\mathrm{eV}$ at 95\% CL. This limit is getting close to $\Sigma m_{\nu,\mathrm{min}}=0.1\,\mathrm{eV}$, the minimal value allowed by neutrino oscillation experiments in the inverted hierarchy scenario. Future cosmological experiments will reach a sensitivity of $\sigma(\Sigma m_\nu)=0.015\,\mathrm{eV}$ and the possibility to discriminate between normal and inverted hierarchy, provided that $\Sigma m_\nu<0.1\,\mathrm{eV}$. Cosmological data are also in agreement with the expectations of the standard cosmological model which predicts a number of relativistic species at early times of $N_\mathrm{eff}=3.045$. Current data constrain $N_\mathrm{eff}=3.15\pm0.23$ at 68\% CL and reject the presence of an additional fully-thermalized species, such as a light sterile neutrino, at $>3\sigma$ level. Future experiments will reach the sensitivity of $\sigma(N_\mathrm{eff})=0.027$. This sensitivity will enable to test at $1\sigma$ level the physics of non-instantaneous neutrino decoupling and the presence of a Goldstone boson decoupling before QCD phase transition.

%
%


\Acknowledgements
The participation to NuPhys2017 and this work are supported by the Vetenskapsr\^adet (Swedish Research Council) through contract No. 638-2013-8993 and the Oskar Klein Centre for Cosmoparticle Physics.

\end{document}